\documentclass{article}
\usepackage{epsfig}
\usepackage{amssymb}
\usepackage{amsmath}
\usepackage{amsfonts}
\newcommand \be{\begin{eqnarray}}
\newcommand \ee{\end{eqnarray}}
\begin{document}
\begin{center}
{\bf Low density Neutron Matter at Finite and Zero Temperatures}\\
\bigskip
\bigskip
H. S. K\"ohler \footnote{e-mail: kohler@physics.arizona.edu} \\
{\em Physics Department, University of Arizona, Tucson, Arizona
85721,USA}\\
\end{center}
\date{\today}

\begin{abstract}
This report concerns the energy of neutron-matter
for densities below $0.15 fm^{-3}$ and temperatures at and below $10 MeV$. 
Separable NN-interactions are obtained by inverse scattering 
from the experimental phase-shifts with specified  momentum cut-offs $\Lambda$.
Results of Brueckner-Bloch-DeDomicis as well as finite temperature Green's 
function calculations show independence of cut-off for $\Lambda \geq 3$.
Agreement with the low-density virial expansion is found.
Results of Hartree-Fock as well as second  order calculations show  
considerable $\Lambda$-dependence and the agreement with the virial
expansion is lost.
The "best" first-order choice of $\Lambda$ is found to be $\sim 2.5
fm^{-1}$, which agrees with $V_{low \ k}$ studies. Reasonable agreement between
the second order and the Brueckner results is also found for this value of 
$\Lambda$ except at low density.  Only 2-body forces are used in this study.
\end{abstract}

\section{Introduction}
The equation of state (EOS) of neutron matter is of particular interest in
astrophysical studies.  The mass of neutron-stars is related to the EOS at
high density $\sim 0.5-0.7 fm^{-3}$.  
Lower densities $\leq 0.15 fm^{-3}$ are however of interest in problems
related to supernova explosions.
The zero temperature Brueckner theory is a well tested method at such
densities. A finite temperature extension of this theory was given by Bloch
and De Domicis.\cite{blo58} It is used here and comparisons are made with
the finite temperature
Green's function method. A main difference
from the Brueckner method is that the latter includes the spectral broadening
self-consistently and the consequences of this broadening has to be
investigated. Also, the spectral function contains 
all one-body properties. In that sense it supersedes the Brueckner method.
It connects seamlessly to the non-equilibrium Green's function
quantum transport theory being the stationary solution of this
time-dependent theory. 
The spectral functions approach a quasi-particle (Brueckner)
limit at low temperatures and low density and the energy-integrations 
that are part of the Green's function method  then become numerically difficult. 
The Green's function method is therefore used only at high density and
temperature and  comparison with the Brueckner-Bloch-DeDomicis method shows
satisfactory agreement here.

There are numerous publications related to the energy of a low
density neutron gas, mostly at zero temperature. The results in one of the
earliest\cite{soo60} already agree satisfactorily with the most 
recent\cite{bal07,tol07,sch05} at densities low enough for the $^1S_0$-state
to be dominant, even though the methods have been different.
The finite temperature equation of state has been studied in detailed
calculations by Margueron et al\cite{mar03} and by Baldo et al\cite{bal98}
using Bloch-De Domicis methods.
It is the focus of the present work. Our study also
relates to recent works by Tolos et al\cite{tol07} using
Kohn-Luttinger-Ward methods, and by Horowitz and Schwenk\cite{hor06}.
A main purpose of the present study is to illustrate the 
usefulness of separable interactions that are directly related to
scattering phase-shifts by inverse scattering  to investigate the
dependence on cut-offs in momentum-space. The study should be
considered as exploratory.
Any "final" precision calculation of
nuclear properties should be based on 2- 3- (and
many-body) forces derived from first principles. 

The potentials are fitted to on-shell scattering data and the ranks of the
potentials are the minimum required for these fits. If off-shell data are
available these can also be fitted by increasing the rank. These data could
in principle come directly from experiments or from meson-theoretical
potentials. In case of the $^3S_1-^3D_1$-states
the  deuteron provides such off-shell information. This was used in
ref\cite{kwo95}. For the problem at hand, neutron-matter, this is not
relevant. Here the main state is the $^1S_0$. For reasons stated below this
state is expected to be a good candidate for a separable interaction. It was
indeed found in ref\cite{kwo95} that the half-off shell reactance matrix
elements reproduced the Bonn-B potential data. 

The basic equations used for the inverse scattering problem, the effective
interaction in Brueckner theory (the G-matrix) and the role of 
the momentum cut-off is presented in Section 2. A short summary of the
Green's function method with its  effective interaction (the in-medium T-matrix) 
is given in Section 3. 
Section 4 presents the results of the numerical calculations while 
a summary of the findings is found in Section 5.

\section{Separable NN-interaction and Brueckner Theory}
Because the $S$-states have poles near $E=0$ a reasonable ansatz is to
assume that these potentials are separable.\cite{bro76}
The inverse scattering method allows the numerical construction of a
separable potential that reproduces a given set of phase-shifts EXACTLY.
This method was used in ref.
\cite{kwo95} with application to the energy of symmetric nuclear matter. 
The agreement with Bonn-B results was particularly impressive for the
$^1S_0$-state but also showed surprising agreement for all states except
the $^3P_1$ for which it was found that the half-shell reactance matrix
differed substantially. The Arndt phases\cite{arn87} were used then and are also
used here but at low energy supplemented by the phases obtained from
the NN scattering length ($a_{nn}=-18.5fm$) and effective range ($r_c=2.68
fm$).
The effect of low-momentum cut-offs was investigated in several
subsequent papers \cite{hsk08,hsk07,hsk06,hsk04} and the connection with 
$V_{low \ k}$ was shown. (See for example Fig. 1 in ref. \cite{hsk04}).
Readers are referred to these earlier publications for details regarding
inverse scattering.
Only a short summary and some important features of the method are shown below.
A rank one separable potential is adequate
at low density with low relative momenta between the nucleons.
But the $^1S_0$ phases turn repulsive above $k_c\sim 1.6 fm^{-1}$ necessitating a
rank 2 potential as a minimum requirement to fit these phase-shifts and
others behaving similarly. Below
are only presented the rank one formalism applicable for momentum cut-offs
$\Lambda \leq k_c$. For larger cut-offs that require a rank 2 potential
the equations are modified as in
previous work \cite{kwo95} using the Bolsterli and MacKenzie
method\cite{bol65}. This is (of course) not a unique method of constructing
a potential but as stated
above it was found to give good agreement with the Bonn and presumably with
other realistic potentials.\cite{kwo95}

For a rank one separable interaction one has, with the cut-off $\Lambda$
indicated explicitly as a parameter
\begin{equation}
<k|V_{\Lambda}||p>=-v(k;\Lambda)v(p;\Lambda)
\label{V}
\end{equation}
From inverse scattering one finds:
\begin{equation}
v^{2}(k;\Lambda)= \frac{(4\pi)^{2}}{k}sin \delta (k)|D(k^{2})|
\label{v2}
\end{equation}
where
\begin{equation}
D(k^{2})=exp\left[\frac{2}{\pi}{\cal P}\int_{0}^{\Lambda}
\frac{k'\delta(k')}{k^{2}-k'^{2}}dk' \right]
\label{D}
\end{equation}
where ${\cal P}$ denotes the principal value  and
$\delta(k)$  is the scattering phase-shift. 

Fig. 1 in ref.\cite{hsk06} shows results of Brueckner calculations 
with this potential in the $1S_0$-state.
There are two curves. The uppermost shows the potential energy as a
function of $\Lambda$ with
dispersion-correction, the lower without this correction.  The density is
given by $k_f=1.35 fm^{-1}$.
The difference between the two curves is small. 
What is also important here is that the energy is 
constant for $\Lambda \geq \sim 2.5 fm^{-1}$. 
It is however equally important that this is a result of a
ladder-summation to all ordersi, Brueckner theory.

The interaction obtained from the inverse
scattering depends on the parameter $\Lambda$ as shown already by Fig. 3 in
ref.\cite{hsk04} so to first order in the interaction the energy is not
expected to be independent of $\Lambda$. This is demonstrated in numerical
examples below where the second and all order terms are also considered.

\begin{figure}
\centerline{
\psfig{figure=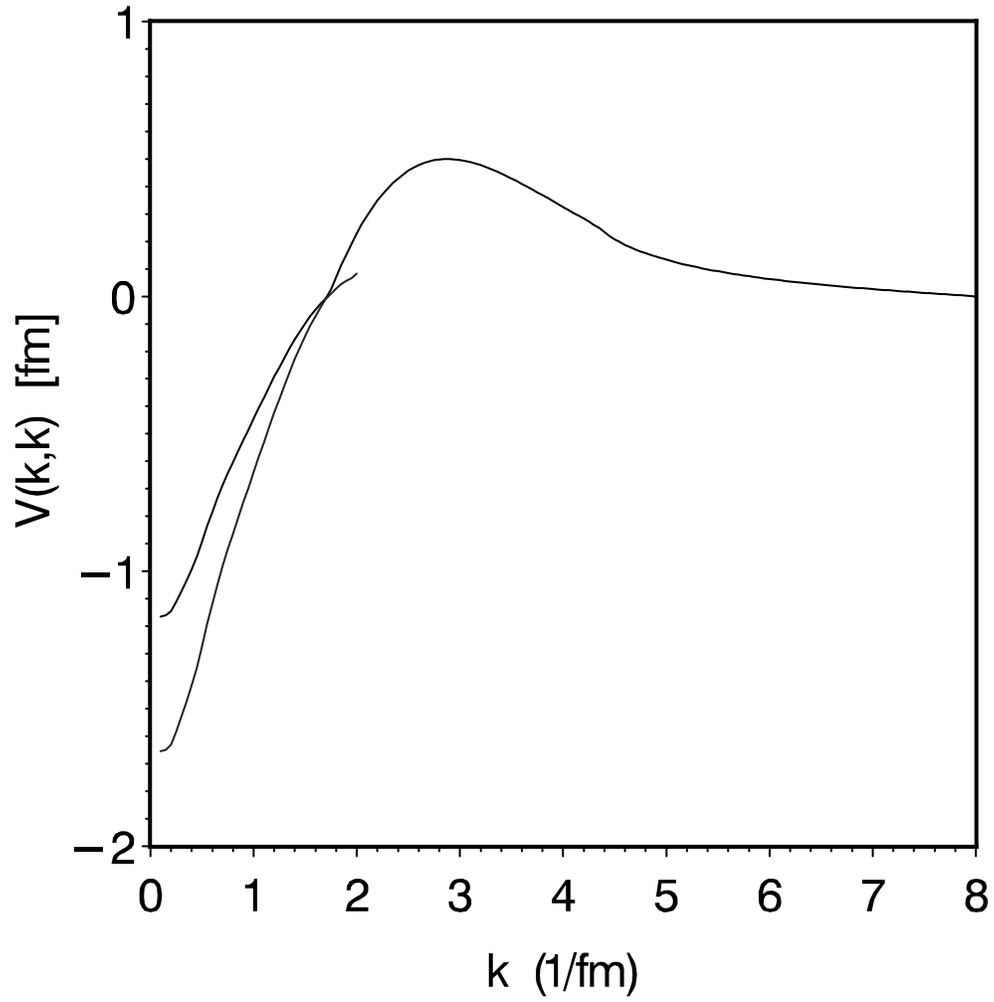,width=15cm,angle=0}
}
\vspace{.0in}
\caption{
These curves illustrate the dependence of the $^1S_0$-interaction on
$\Lambda$ as obtained from the inverse sacttering. 
The maximum value of momentum for each of the two curves is at
the two chosen values of $\Lambda$, $2$ and $8 fm^{-1}$ respectively.
}
\label{vboss}
\end{figure}

The $\Lambda$-dependence is further illustrated here by
Fig. \ref{vboss} showing $<k|V_{\Lambda}|k>$ for two values of $\Lambda$.
Although the two potentials are notably very different from each other they
have in common that they both fit the phase-shifts EXACTLY for all $k<2
fm^{-1}$. The $\Lambda=8 fm^{-1}$ potential of course ALSO fits the phases from
$2\rightarrow 8 fm^{-1}$.

This means that the diagonal of the reactance matrix defined by 
\begin{equation}
K(k)=-\frac{v^{2}(k;\Lambda)}{1+ I_{K}(k)}=-4\pi\cdot tan\delta(k)/k
\label{K}
\end{equation}
with
\begin{equation}
I_{K}(k)=\frac{1}{(2\pi)^{3}}{\cal P}\int_{0}^{\Lambda}
\frac{v^{2}(k';\Lambda)}{k^{2}-k'^{2}}
k'^{2}dk'
\label{I_K}
\end{equation}
is  independent of $\Lambda$ for $k\leq \Lambda$. 
Notice that eq. \ref{K} implies a summation to all orders in $V$.
The in-medium interaction defined by the Brueckner $G$-matrix
(or by the in-medium  $T$-matrix defined with Green's functions) differs
from the reactance matrix $K$ by the Pauli-blocking. 
Neglecting the dispersion-correction which in our case is small (see
above),
i.e. with no self-energy in the propagator, we have
\begin{equation}
 G(k,P)=-\frac{v^{2}(k)}{1+ I_{G}(k,P)}
\label{G}
\end{equation}
with
\begin{equation}
I_{G}(k,P)=\frac{1}{(2\pi)^{3}}\int_{0}^{\Lambda}
v^{2}(k')\frac{Q(k',P)}{k^{2}-k'^{2}}
k'^{2}dk'
\label{I_G}
\end{equation}
where $P$ is the center of mass momentum and  $Q$ the angle-averaged
Pauli-operator for pp-ladders, being exact with only kinetic energy in the
denominator. 

\begin{figure}
\centerline{
\psfig{figure=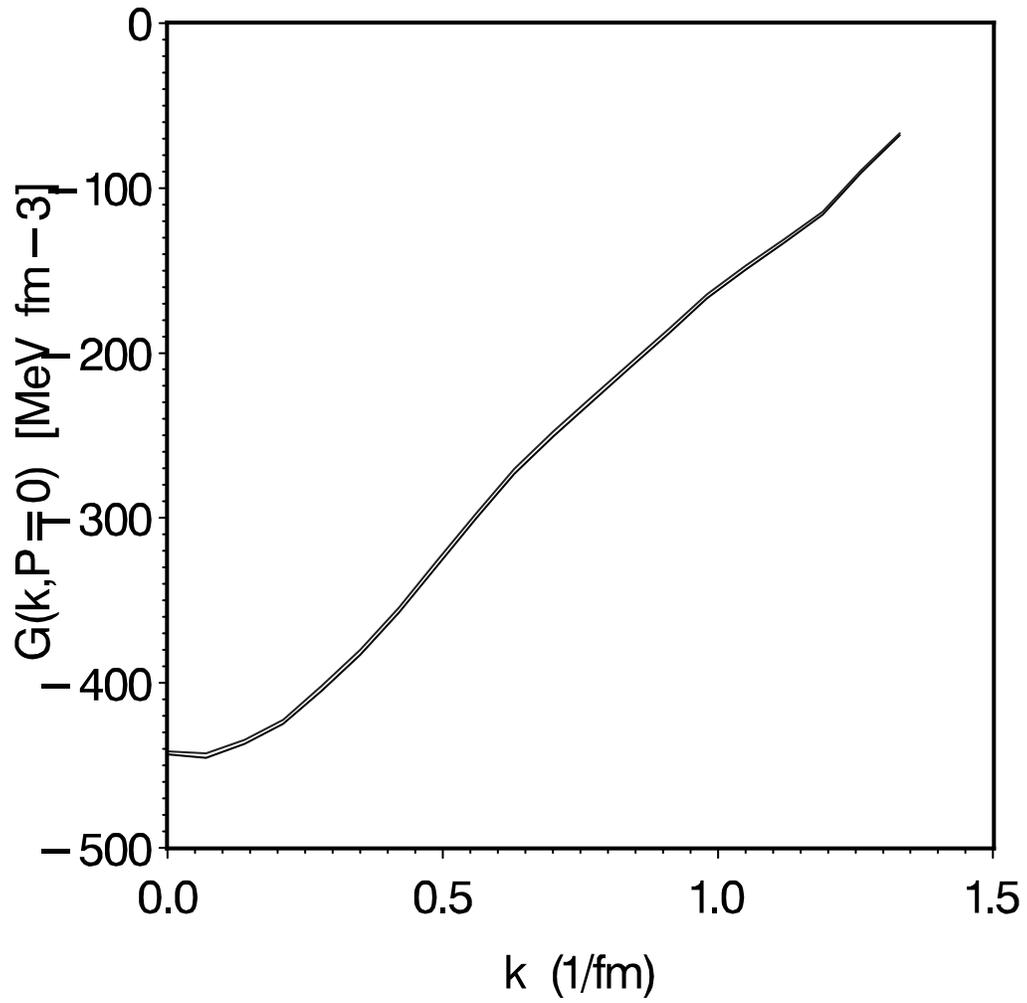,width=15cm,angle=0}
}
\vspace{.0in}
\caption{
$G(k,P=0)$ calculated from the two potentials of Fig. \ref{vboss}.
There is complete overlap of the two curves. The fermi-momentum is here
$k_f=1.0 fm^{-1}$.
}
\label{aff}
\end{figure}

With the Pauli-blocking one has to expect $G$ unlike $K$  to be
$\Lambda$-dependent. Calculations show however near independence even here as
long as $\Lambda\geq 2k_f$.\cite{hsk06} 
This is exemplfied by Fig. \ref{aff} showing $G$-matrix elements 
calculated from the two potentials
of Fig. \ref{vboss}. The two curves overlap completely even though the
potential are quite different. The center-of-mass
momentum is here chosen to be zero, which  gives the maximum effect of the
$Q$-operator.

Further numerical investigations of the $\Lambda$-dependence as well as 
of ladder-expansions will be shown below. A few observations can be 
done without reference to detailed calculations. 
Let the density of nucleons $\rightarrow 0$. Then $Q\rightarrow 1$. 
It is interesting to observe that in this limit
$$G\rightarrow -4\pi\cdot\delta(k)/k,$$ referred to as the 
phase-shift approximation \cite{rie56} with no dependence on
$\Lambda$. (Note the limit gives $\delta$ NOT $tan\delta$).
But it is imporant here that this requires  a complete (all order)
ladder-summation to be made as in eq. (\ref{G}).  

One might expect the  phase-shift approximation to be
useful for a low-density neutron-gas. It was however already found  by Sood
and Moszkowski \cite{soo60}  that a proper treatment of
the Pauli-blocking at low density, ($0.1<k_f<0.5 fm^{-1}$) and zero
temperature
reduces the potential energy/neutron by a factor of $\sim 1/2$ relative to
this approximation. The dimensionless quantity to consider is in fact
$k_f\cdot a_s$. With $a_s \sim 20 fm$ for the neutrons this would 
require $k_f\ll .05$, i.e.
an extremely low density for the phase-shift approximation to be
valid.\footnote{I thank Prof. Nai Kwong for helpful discussions relating to
this observation.}

A first order approximation $G(k)\sim<k|V_{\Lambda}|k>$ would on the other
hand be valid only for complete blocking, or for a weak potential in which
case $G\sim V \sim \delta/k$. (With $\Lambda$ sufficiently small, i.e. 
$V_{low \ k}$ weak enough, one may hope to reach this limit.)

Blocking also increases with density so that a first order approximation
may become useful in this case.
This is consistent with the findings by Moszkowski and
Scott.\cite{mos60} They show the second order  contribution 
from their long-ranged part to decrease with increasing density.
The long-ranged part in their separation-method  corresponds to our
low-momentum interaction with $\Lambda\sim 2-3fm^{-1}$ and 
it also relates to $V_{low \ k}$.\cite{hol04} 
Numerical agreement between the Moszkowski-Scott method and $V_{low \ k}$ was
shown in ref. \cite{hsk04}.

It was already emphasised above that eq. (\ref{G}) for the $G$-matrix is a ladder 
summation to all orders of the bare interaction.  Below we shall also 
do a calculation to second order using $G\rightarrow G^{(2)}$ with

\begin{equation}
<k|G^{(2)}|(P,\omega)|p>=v(k)v(p)(1-I_G(P,\omega))
\label{G2}
\end{equation}

Note that with a separable interaction there is no difference in 
computing effort beteween second and all orders.

Calculations will also be shown below to first order i.e. with

\begin{equation}
<k|G^{(1)}|(P,\omega)|p>=v(k)v(p)
\label{G1}
\end{equation}

All Brueckner calculations were done 
neglecting the dispersion correction, i.e. without self-energy insertions 
in  particle- and in hole-lines. This approximation is justified by the
results of ref. \cite{hsk06} showing this correction to be small in the
$^1S_0$ channel. (See also  ref.\cite{bal07}.) But it is not
part of the Green's function calculations. There, the calculation of the
$T$-matrix involve integrations over the spectral-functions that by their
definition implicitly contain the self-energies.

In the Brueckner case a combination of eqs (\ref{G}) and (\ref{K}) 
gives \cite{hsk07,hsk08}.
\begin{equation}
G(k,P)=-\frac{v^{2}(k;\Lambda)}{I_{GK}(k,P)}
\label{GK}
\end{equation}
with
\begin{equation}
I_{GK}(k,P)=\frac{1}{(2\pi)^{3}}\int_{0}^{2k_f}
v^{2}(k';\Lambda)\frac{Q(k',P)-{\cal P}}{k^{2}-k'^{2}}
k'^{2}dk'+\frac{kv^{2}(k;\Lambda)}{\tan\delta(k)}
\label{IGK}
\end{equation}

Eqs (\ref{GK},  \ref{IGK}) have the advantage over eqs (\ref{G},
\ref{I_G})
in that the integrand in eq. (\ref{IGK}) is zero for  $k'>2k_f$
because   the factor $Q(k',P)-\cal{P}$ is then equal to zero.
Although the two sets of equations are numerically
identical the latter set simplifies the computing greatly.
The high momentum component of the interaction is eliminated. This new
equation also shows that $G$ is independent of scaling $v(k;\Lambda)$ with
a constant factor. Fig. \ref{vboss} shows this to be approximately true and
it at least partly explains the result shown by Fig. \ref{aff}, the
independence of $\Lambda$. 
Note however that the above eqs (\ref{GK}) and (\ref{IGK}) are obtained
only if assuming the neglect of the dispersion correction mentioned above.

Apart from this modified expression, used here at zero temperature
the Brueckner calculation of
the energy proceeds as described repeatedly in the literature. 

At finite temperatures the Pauli-blocking is modified as shown by Bloch
and DeDomicis.
Only two-body interactions were included here. Obviously, 3-body etc and
higher order graphs become important with increasing density.

\section{Green's Function Method}
Expressions and discussions relating to the Green's function equations 
for the in-medium T-matrix, self-energies and total energy of a
fermion-system can be found in numerous publications.
\cite{hsk93,boz02,boze02,fri03}
Only a summary of equations used in this work are shown here.

Like in the previous section the formalism is only shown for a rank one
potential with the modifications for the rank 2 done as in ref.\cite{kwo95}.

With the separable two-body interaction defined above 
the  in medium $T$-matrix  is given by (the dependence on $\Lambda$ is
suppressed but is implicit)
\begin{equation}
<k|T|(P,\omega)|p>=\frac{v(k)v(p)}{1-I(P,\omega)}
\label{T}
\end{equation}
where $P$ is the ceneter of mass momentum and
\begin{equation}
I(P,\omega)=\int \frac{k^2v^2(k)dk}{(2\pi)^3}\int dcos(\theta)\int
\frac{d\omega'}{2\pi}\int\frac{d\omega''}{2\pi} \nonumber \\
\frac{S(p_1,\omega'-\omega'')S(p_2,\omega'')(1-f(\omega'-\omega'') -f(\omega''))}
{\omega-\omega'+i\eta}.
\label{I}
\end{equation}
where $$p_{1,2}=\frac{P^2}{4}+k^2\pm Pkcos(\theta).$$

The imaginary part of the selfenergy is given by
\begin{equation}
Im\Sigma(p,\omega)=\frac{3}{8\pi}\int\frac{d\omega'}{2\pi}\int\frac{{\bf
dk}}{(2\pi)^3} \nonumber \\
<\frac{{\bf p-k}}{2}|ImT(|{\bf p+k}|,\omega+\omega')|\frac{{\bf
p-k}}{2}>S(k,\omega')(f(\omega')+b(\omega+\omega')).
\label{imsig}
\end{equation}
where $b(\omega)$ is the Bose function.
The real part $Re\Sigma$ is obtained by the usual dipersion relation 
and the HF-term is added.

With the self-energies given, the spectral function is calculated and a new
T-matrix is obtained. The system of equations are iterated until
convergence. At each iteration the chemical potential is calculated from
the nucleon density. 
The correlated distribution function differs from the fermi- distribution
and is given by
\begin{equation}
n(p)=\frac{1}{2\pi}\int f(\omega)S(\omega,p)d\omega
\label{dis}
\end{equation}
and the removal energy is
\begin{equation}
r(p)=\frac{1}{2\pi n(p)}\int \omega f(\omega)S(\omega,p)d\omega
\label{rem}
\end{equation}

The total energy is given by 
\begin{equation}
E= \frac{1}{(2\pi)^3}\int d{\bf p} [p^2/2m-r(p)]n(p)
\label{etot}
\end{equation}

\section{Numerical Results}

\subsection{Energy-Density Results;$^1S_0$.}
\begin{figure}
\centerline{
\psfig{figure=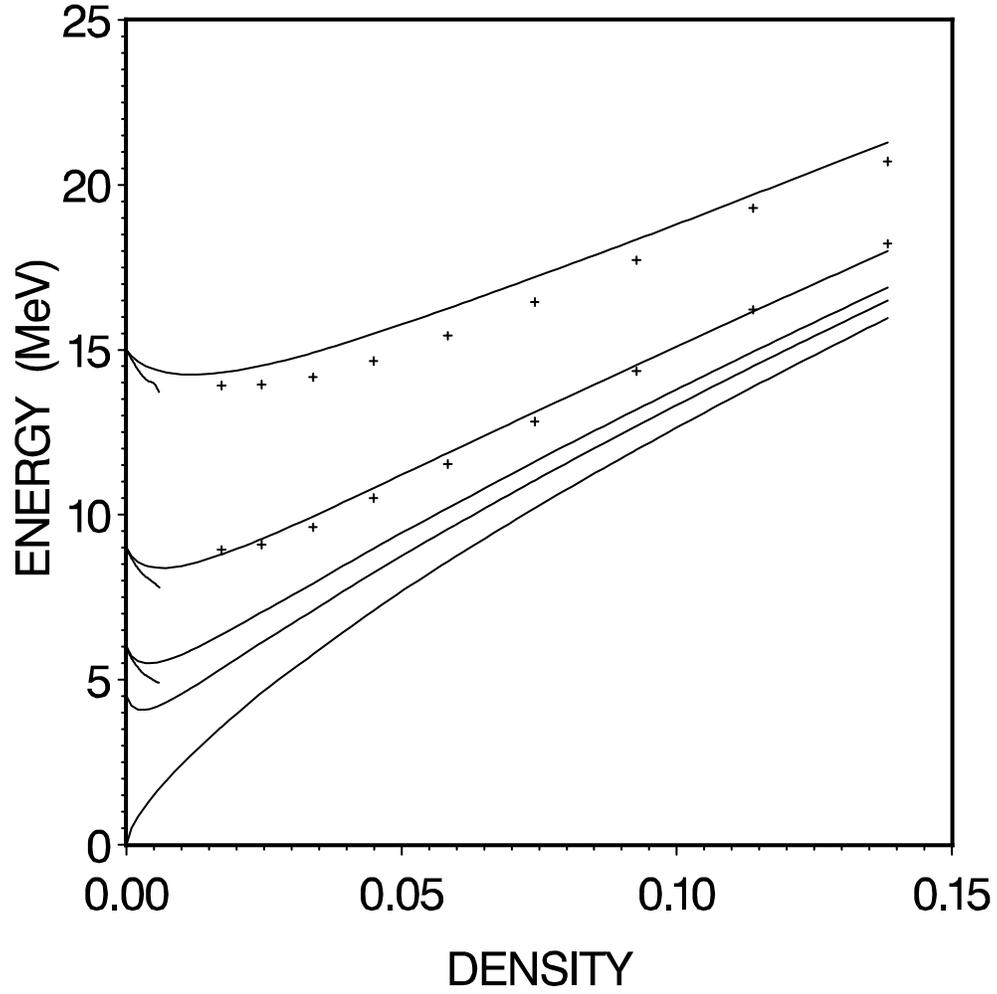,width=15cm,angle=0}
}
\vspace{.0in}
\caption{
The solid lines show the energy at zero, $0,3,4,6$ and $10 MeV$
temperatures respectively from Brueckner ladder summations. The crosses are
from Green's function calculations with selfconsistent spectral functions
and with in-medium $T$-matrix ladders to all orders defined by eq. (\ref{T}).
Only $^1S_0$-states included here.
The short lines are low-density virial results from ref.\cite{hor06}.
}
\label{neutronbr}
\end{figure}

Fig. \ref{neutronbr} shows the energy per particle of the neutron-gas as a
function of density at temperatures $0,3,4 ,6$ and $10 MeV$. The full lines
are from the Brueckner-Bloch-DeDomicis calculations. The crosses are from
the Green's function calculations. The short full lines are virial
results  obtained from the work of Horowitz and Schwenk \cite{hor06}.

The Brueckner calculations were done using standard methods with the
cut-off $\Lambda=3 fm^{-1}$ and/or $\Lambda=6 fm^{-1}$ at the highest 
density. Increasing $\Lambda$ did not change the results while  a decrease
of $\Lambda$ showed noticable differences as shown below in section 4.2. 
It can be remarked that the
Brueckner calculations with the separable potential and inverse scattering are
extremely simple and flexible. Any set of inputted phase-shift data
provides an output of binding energy in seconds. Dependences on $\Lambda$
are easily explored.
The momentum-mesh was $0.01 fm^{-1}$. 

For the Green's function calculations, the set of equations (in medium T-matrix,
self-energy $\Sigma 's$ and spectral functions) were iterated until the 
total energy and chemical potential were stationary. This could require 
anywhere from 5 to 20 iterations depending on the input spectral functions. 
The in medium T-matrix was calculated with the separable potential as defined
above and with the momentum cut-off kept constant at
$\Lambda=3 fm^{-1}$. The  $\omega$-integrations were typically from $-250$ to
$+250 MeV$  with a mesh  of $\sim 1 MeV$.

The Green's function calculations are quite lengthy compared with the
Brueckner calculations. They require a factor of thousands more computer-time. 
Although the equations (and the calculations) are rather
different in the two cases, the results are surprisingly similar. This seems
to agree (qualitatively) with zero-temperature results of other authors.
\cite{boz02,fri03}
A bonus is that the spectral functions contain more information
on one particle properties such as distribution functions and removal energies.
Using Thouless criterion\cite{tho60}
allows a determination of the BCS critical
temperature. This was not investigated in detail but it was determined that
$T_c\sim 3 MeV$ at $n\sim 0.1 fm^{-3}$.
The spectral functions approach a quasi-particle
limit at low temperatures and low density (See Fig. \ref{spec8} below) 
and the energy-integrations then become numerically difficult. 
This can be resolved by resorting to the
Extended Quasiparticle Approximation (EQP)\cite{hsk93,lip01}
or a similar method used by B\.ozek \cite{boze02}. This was not done here
as the Brueckner-Bloch-Domicis method appeared adequate for the present 
and also numerically so much easier.

\subsection{$\Lambda$-dependence and low-order expansion;$^1S_0$}
The dependence on the cut-off in momentum-space was discussed above. Here
we show numerical results both for the complete Brueckner ladder summation,  
eq. (\ref{G}), as well as for the second and first order expansions, 
eqs (\ref{G2}) and (\ref{G1}).

\begin{figure}
\centerline{
\psfig{figure=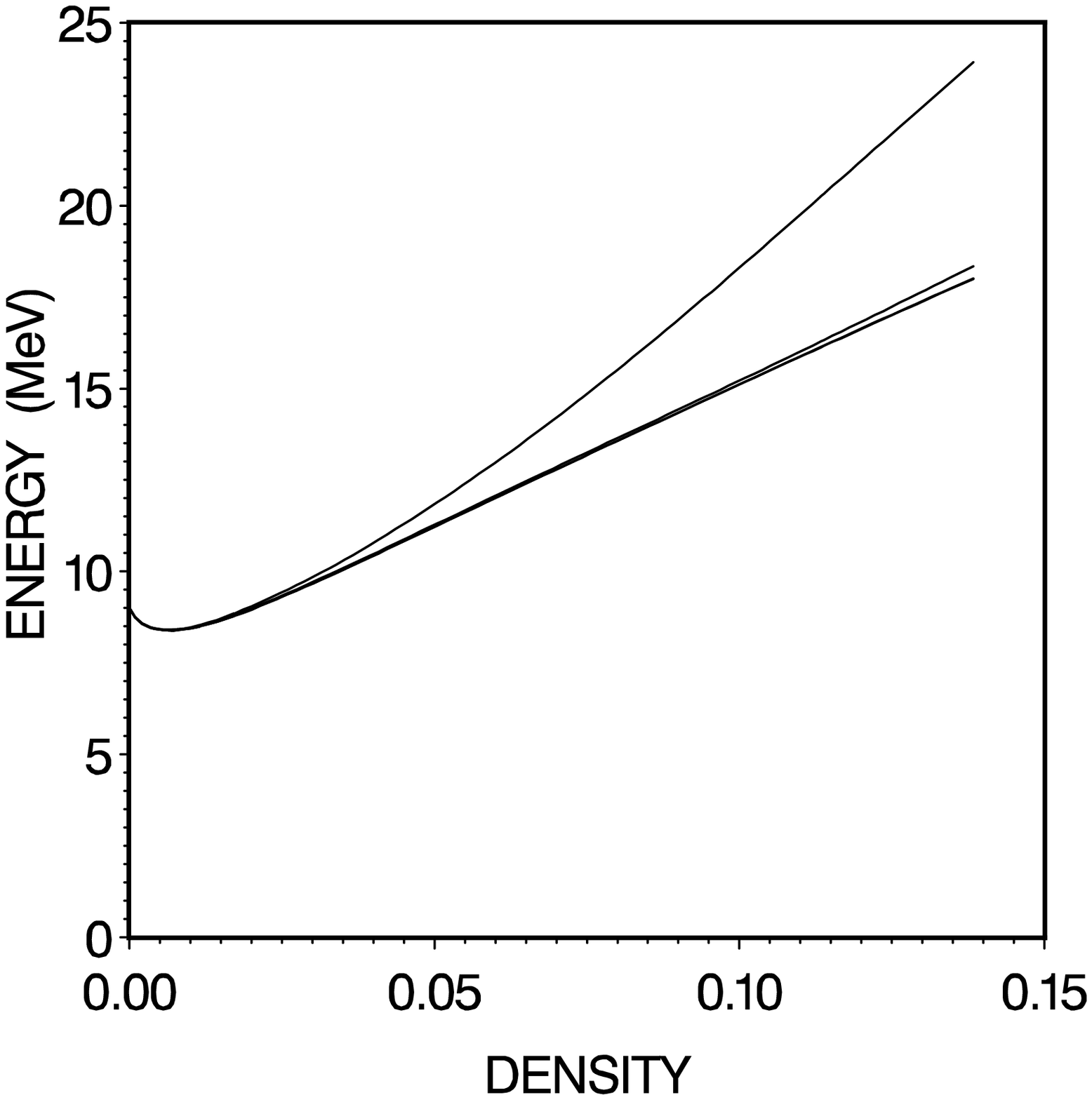,width=15cm,angle=0}
}
\vspace{.0in}
\caption{
Shown are full $G$-matrix results, eq.(\ref{G}), at $T=6 MeV$. The uppermost
curve is for $\Lambda=2 fm^{-1}$, followed by $\Lambda=3,6$ and $8
fm^{-1}$, the latter completely overlapping. $^1S_0$-states only.
}
\label{nbrlambda}
\end{figure}

Fig. \ref{nbrlambda} shows the energy per particle at $T=6 MeV$ with the full
$G$-matrix, eq. (\ref{G}). Calculations were done for $\Lambda=2,3, 6$ and
$8 fm^{-1}$ with the Figure showing all results converging at the lowest
densities. At higher densities the $\Lambda=2$ curve is appreciably higher
while for $\Lambda\geq 3$ one sees convergence.

\begin{figure}
\centerline{
\psfig{figure=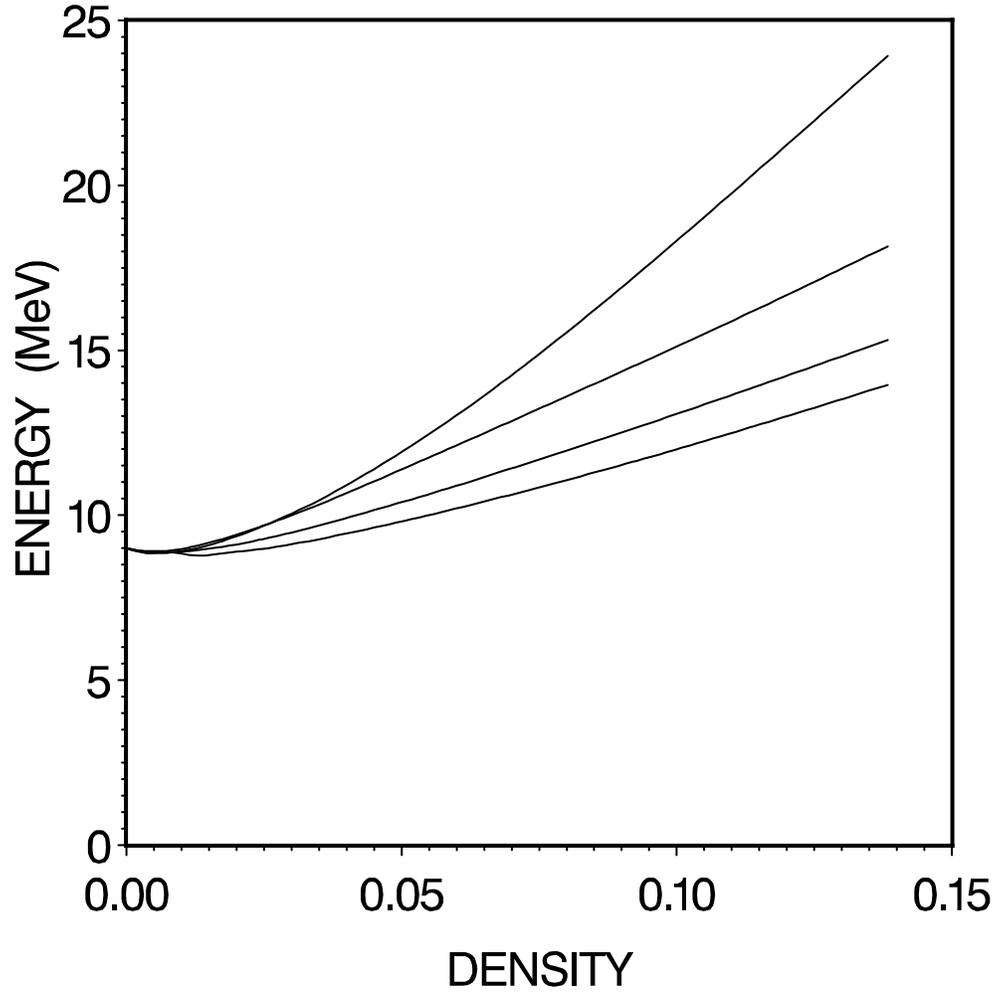,width=15cm,angle=0}
}
\vspace{.0in}
\caption{
Shown are second order results eq.(\ref{G2}) at $T=6 MeV$. The uppermost
curve is for $\Lambda=2 fm^{-1}$, followed by $\Lambda=3,6$ and $8
fm^{-1}$. $^1S_0$-states only.}
\label{n2lambda}
\end{figure}

\begin{figure}
\centerline{
\psfig{figure=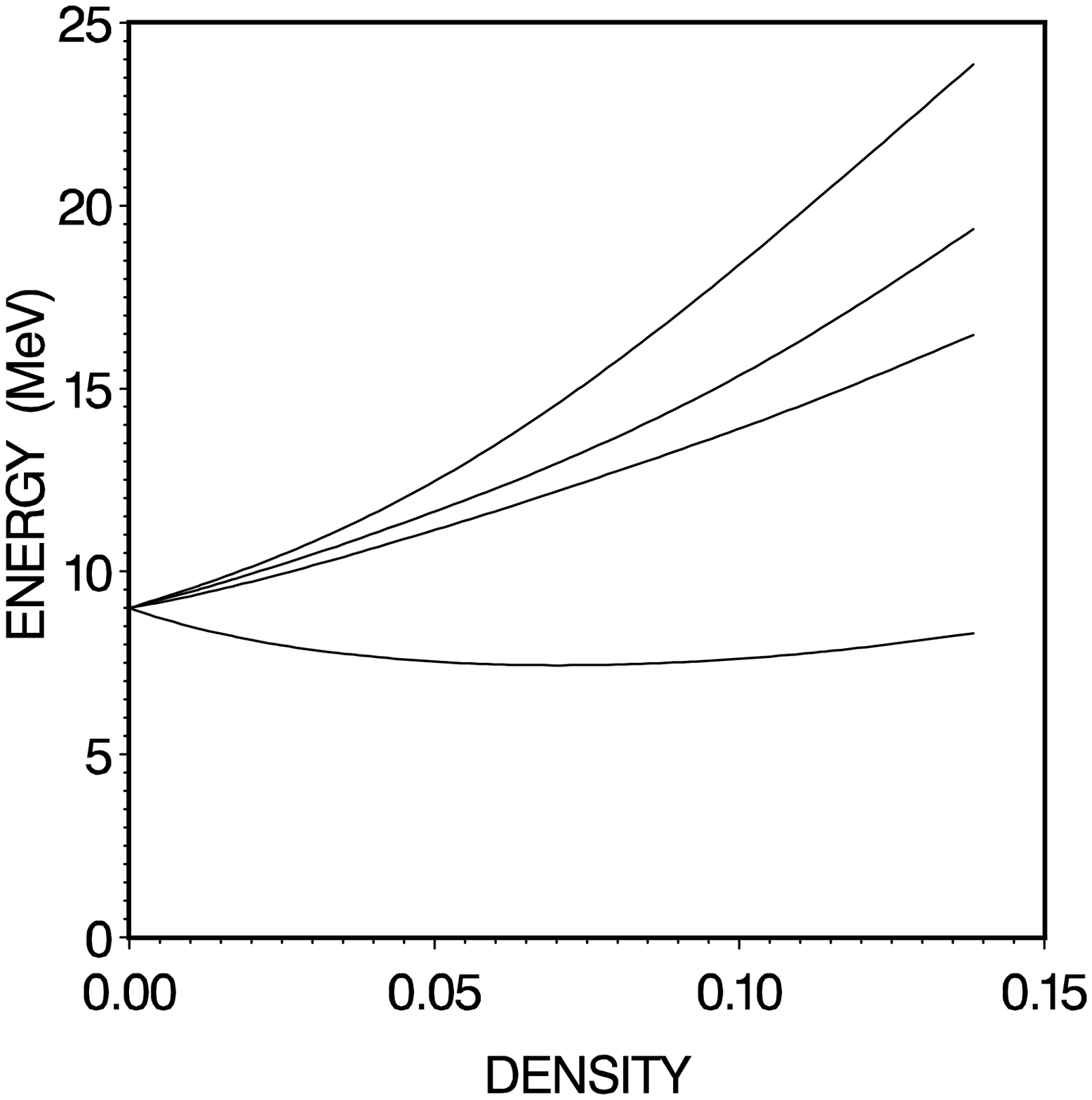,width=15cm,angle=0}
}
\vspace{.0in}
\caption{
Shown are first order calculations eq.(\ref{G1}) at $T=6 MeV$. The uppermost
curve is for $\Lambda=2 fm^{-1}$, followed by $\Lambda=2.5,3$ and $6
fm^{-1}$.} Note that the $\Lambda=8$ curve is not shown. It would be much
below the others. $^1S_0$-states only.
\label{n1lambda}
\end{figure}

Fig. \ref{n2lambda} shows second order results, using eq. (\ref{G2}). 
They differ appreciably from the full  $G$-matrix calculations as does the
first order results shown in Fig. \ref{n1lambda}.
The preferred $\Lambda$ showing the best over-all agreement with the full
$G$-matrix is $\Lambda \sim 2.5$ for $G^{(1)}$ and $\Lambda \sim 3$ for
$G^{(2)}$. Comparing Fig. \ref{nbrlambda} with Fig. \ref{n2lambda} one sees
almost perfect agreement between the second order and full $G$-matrix
results both for $\Lambda=2$ and $3 fm^{-1}$. Comparison with the first
order results of Fig. \ref{n1lambda} shows also good agreement there  for
$\Lambda=2$ but less  so for $\Lambda=3 fm^{-1}$. 
The force weakens if $\Lambda$ decreases so that higher order terms then
of course become less important. 
This is in qualitative agreement with the renormalisation  results leading to
$V_{low \ k}$.\cite{tol07}

But only the full $G$-matrix result is capable of showing agreement with
the virial expansion, at low density.

\subsection{Energy-Density;including $L\leq 7$}
\begin{figure}
\centerline{
\psfig{figure=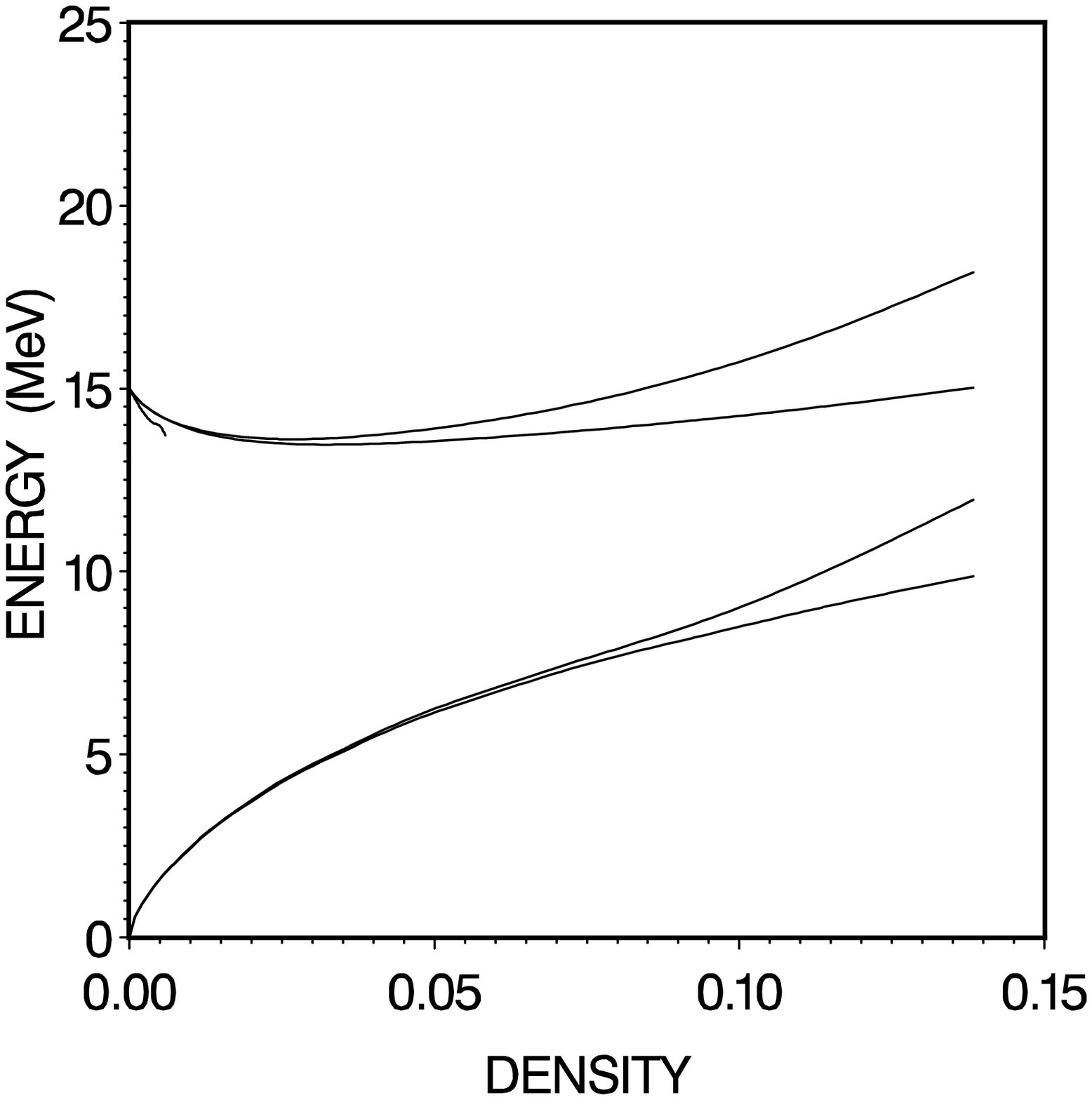,width=15cm,angle=0}
}
\vspace{.0in}
\caption{
Energy as a function of density including all tw0-body states with $L\leq
7$. 
The two top curves are for $T=10 MeV$ with $\Lambda=2.5 fm^{-1}$ 
(upper curve) and $\Lambda=5 fm^{-1}$ respectively. 
The two lower curves are for $T=0$ with the same values of $\Lambda$.
The short line at $T=10 MeV$ is the virial expansion.
}
\label{neutronall}
\end{figure}

Only the $^1S_0$-states were included in the results presented above.
At zero temperature that is sufficient for densities  $\rho \leq
0.03fm^{-3}$.(see e.g.ref.\cite{sch05}). At higher temperatures even the zero
density virial expansion gets contribution from the higher partial waves
\cite{hor06} and 3N-forces are increasingly important with density and
temperature increases.
Fig.\ref{neutronall} shows results including all partial waves $L\leq 7$.
Comparison with Fig. \ref{neutronbr} shows the decrease in energy 
from higher partial waves increasing with density while it is known
that the contribution from 3N forces is repulsive and also
increasing with density with the two effects partially cancelling each other. 
Comparison with Fig. \ref{neutronbr} also confirms the statement made above
regarding the contribution of higher partial waves to the virial expansion.
Fig.\ref{neutronall} also shows the $\Lambda$-dependence to increase
with temperature.

The $T=0$ result for $\Lambda=2.5 fm^{-1}$ agrees well with the NN-only 
results shown in refs\cite{tol07,sch03}.

Second order calculations were also done for $\Lambda=2.5$ with almost
perfect agreement with the full Brueckner as is to be expected from the
results shown in section 4.2. 

\subsection{Spectral Functions etc}
\begin{figure}
\centerline{
\psfig{figure=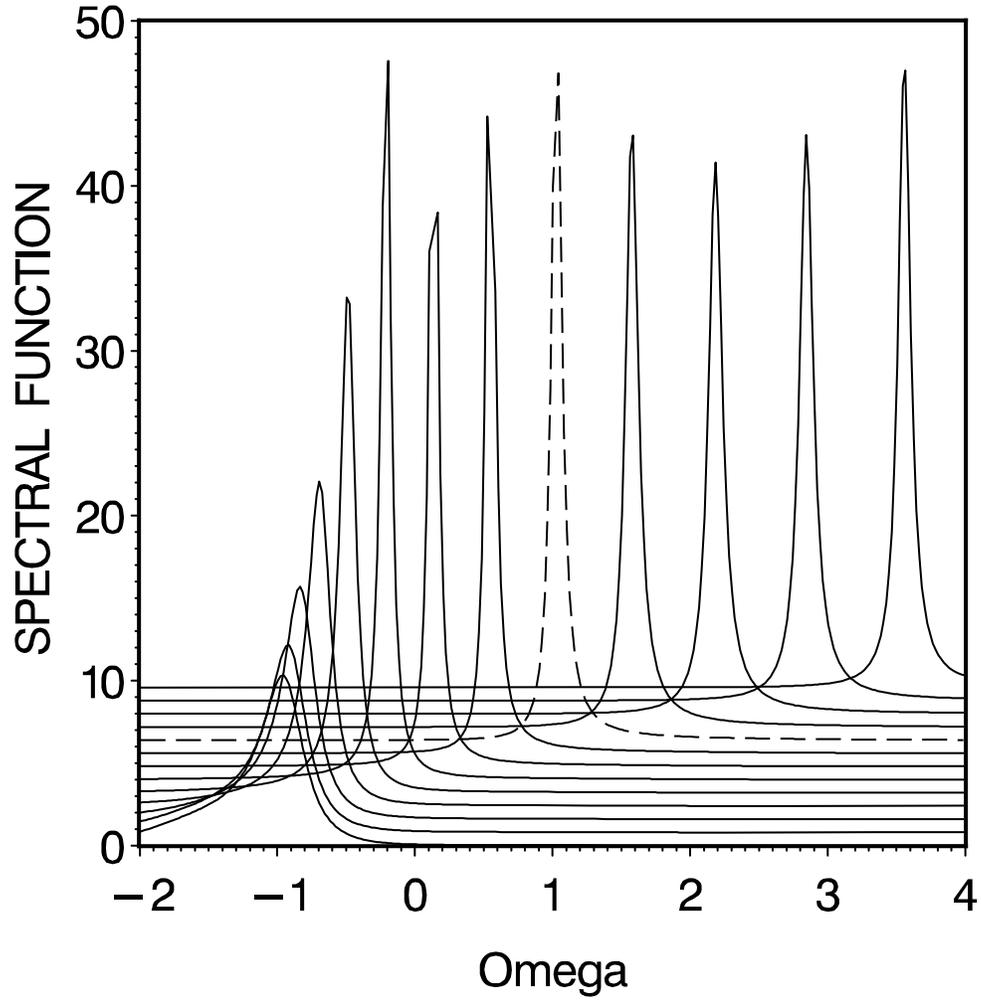,width=15cm,angle=0}
}
\vspace{.0in}
\caption{
Spectral functions $S(p,\omega)$ at $T=10 MeV$ temperature 
and a density of $0.11 fm^3$.
The energy ($\omega-\mu=$Omega) is in units of $\hbar^2/m$.($\mu$ is
chemical potential). The momenta are
offset and range from $p=0$ to the cut-off $\Lambda=3 fm^{-1}$. 
The broken curve is at the fermi-surface. Compare this with Fig. \ref{spec8}
which is for  a lower density. $^1S_0$-states only.
}
\label{spec15}
\end{figure}

\begin{figure}
\centerline{
\psfig{figure=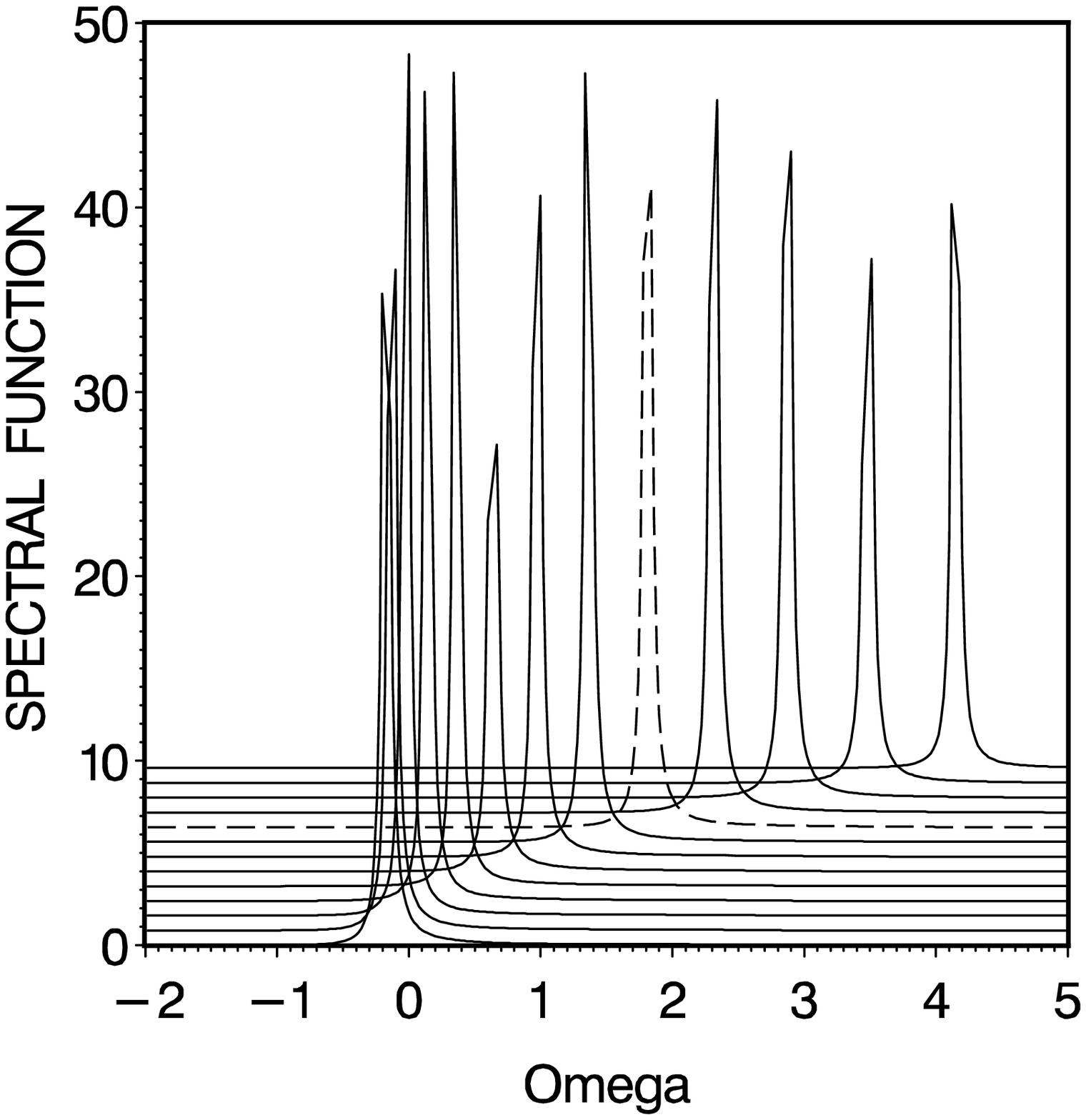,width=15cm,angle=0}
}
\vspace{.0in}
\caption{
Similar to Fig. \ref{spec15} but at a density of $0.017 fm^3$. Notice the
sharply peaked functions indicating the closeness to a quasi-particle limit.
This decreses the accuracy of the $\omega$-integrations  and requires special
treatment. $^1S_0$-states only.
}
\label{spec8}
\end{figure}

The output from Green's function calculations give a more detailed
information about the many-body system than the corresponding Brueckner 
quasi-particle calculations. The additional information is contained in
the spectral functions. Fig. \ref{neutronbr} shows however that the quantity of
interest here, the energy per particle as a function of density, is
practically the same from the Brueckner as from the Green's calculations. 
The Green's function calculations are however computationally much more
demanding, in particular at low density and low temperature where the 
quasiparticle picture dominates.
The figures \ref{spec15} and \ref{spec8} both show spectral functions at a
temperature $T=10MeV$ but the first is at a high density while
the second is at a low density. One may note that these spectral 
functions for the neutron-gas where the two-body interactions are 
only in the $^1S_0$ channel are rather
different from those also including the $^3S_1$ states with stronger
correlations.

\section{Summary}
The energy-density relation for neutron matter at densities ranging from
zero to  $\sim 0.14 fm^{-3}$ and at temperatures ranging from zero to $10
Mev$ have been computed.  
Two-body interactions were obtained from the free scattering 
phase-shifts by inverse
scattering with  momentum-cut-offs  of $\Lambda=2\rightarrow 8 fm^{-1}$. 
Three-body interactions were not included. The many-body 
calculations were made using Brueckners technique
extended to finite temperature by Bloch and DeDomicis. Insertions in hole- and
particle-lines, i.e. dispersion corrections were neglected in this
exploratory study as they are
small for the states and densities considered here. They should of course
be included in a "precision" calculation.

The free scattering interaction is in general
$\Lambda$-dependent as illustrated in Fig.\ref{vboss} while the 
in-medium Brueckner G-matrix in Fig. \ref{aff} shows less or no dependence.
This was discussed and partly explained above in section 2.
Fig. \ref{nbrlambda} shows however that when calculating the total energy
there is a significant difference between the $\Lambda=2$ result and the
$\Lambda \geq 3 fm^{-1}$ results. But there is a nice convergence with
increased $\Lambda$.

Our second order result is noticably $\Lambda$-dependent although less so
than the first order with no sign of convergence in any case and
with considerable difference from the Brueckner results.
The value of $\Lambda$ with the best overall agreement
with the Brueckner results is $\sim 2.5 fm^{-1}$, which seems to agree
with the $V_{low \ k}$ findings. It was also confirmed that for such a
small value of $\Lambda$ second order and Brueckner agree closely except,
as remarked below, at low density..

We note that there is also a close agreement 
with Schwenk and Pethick. \cite{sch05} for $T=0$.

It was found, as did  Tolos et al\cite{tol07}  
that the higher order terms (in $V$) are necessary to
correctly obtain the virial limits at finite temperatures 
associated with the energy minima in Fig.
\ref{neutronbr}. In fact, it was found that the full Brueckner summation
was desirable. But only $^1S_0$ states are included in this figure.
Horowitz and Schwenk\cite{hor06} find that the virial coefficient $b_2$
increases if higher partial waves are also included, thus further 
enhancing the minima.
This increase in $b_2$ becomes larger with increased temperature 
(Table 1. of ref.\cite{hor06}) and is about $20\%$ at the highest 
temperature $T=10 MeV$ in our Fig. \ref{neutronbr}. The virial corrections
shown in this figure  are extracted from ref.\cite{hor06} and 
include all partial waves. The importance of the higher partial waves in
this case is in (qualitative) agreement with Fig.\ref{neutronall}.
This is quite different from the situation at $T=0$ as shown comparing Figs
\ref{neutronall} and \ref{neutronbr}. The $^1S_0$ clearly dominates at low
temperature and density.

Comparisons of the Brueckner results with Green's function results 
were made at $T=6$ and $10 MeV$ and an agreement within numerical 
uncertainties was seen. 

A comparison between Brueckner and Green's function calculations for the
neutron gas at zero temperature was done by B\.ozek et al\cite{boz02}. 
These authors found agreement
between the two methods for the densities considered here and  their
results also agree with the present work. At higher densities they found
the Green's function results much more repulsive than the Brueckner.

The separable interaction used here may be considered
phenomenological. It fits experimental on-shell scattering data. Off-shell
data are not available. As illustarted in ref. \cite{kwo95} it does
reproduce the half-off-shell data of the meson theoretically derived
Bonn-potentials. As such it may be considered "realistic" for the
$^1S_0$-state. The separable
nature of this interaction is supported by the large scattering length
assocated with the nearly bound state in the $^1S_0$ channel.\cite{bro76}
Further comparison with the Bonn-potentials  for 
other states was found in ref.\cite{kwo95}, an
exception being the $^3P_1$-state, requiring in this case that the rank
of the potential be increased.
 
The 3N forces were not considered in the present work. They are important
and should of course be included in the determination of the EOS, but are
also less well determined than the 2N forces. 
A main purpose of the present study was to shed some light on the dependence
on momentum cut-offs and to show that separable potentials derived for
inverse scattering are useful for such a study.

\end{document}